# Attojoule-Efficient Graphene Optical Modulators

Rubab Amin[1], Zhizhen Ma[1], Rishi Maiti[1], Sikandar Khan[1],
Jacob B. Khurgin[2], Hamed Dalir[3], Volker J. Sorger[1,*]

[1]*Department of Electrical and Computer Engineering, George Washington University, Washington, DC 20052, USA*
[2]*Department of Electrical and Computer Engineering, Johns Hopkins University, Baltimore, Maryland 21218, USA*
[3]*Omega Optics, Inc., Austin, Texas 78757, USA*
*Corresponding Author E-mail: sorger@gwu.edu

## Abstract

**Electro-optic modulation is a technology-relevant function for signal keying, beam steering, or neuromorphic computing through providing the nonlinear activation function of a perceptron. With silicon-based modulators being bulky and inefficient, we here discuss graphene-based devices heterogeneously integrated. This study provides a critical and encompassing discussing of the physics and performance of graphene modulators rather than collecting relevant published work. We provide a holistic analysis of the underlying physics of modulators including the graphene's index tunability, the underlying optical mode, and discuss resulting performance vectors of this novel class of hybrid modulators. Our results show that the reducing the modal area, and reducing the effective broadening of the active material are key to improving device performance defined by the ratio of energy-bandwidth and footprint. We further show how the waveguide's polarization must be in-plane with graphene such as given by plasmonic-slot structures. A high device performance can be obtained by introducing multi- or bi-layer graphene modulator designs. Lastly, we present recent results of a graphene-based hybrid-photon-plasmon modulator on a silicon platform, requiring near Boltzmann approximation (100mV) low drive voltages. Being physically compact this 100 aJ/bit modulator opens the path towards a new class of attojoule opto-electronics.**

## 1. Introduction

The functionality of optical modulation bears versatility in both optical communication and information processing [1,2]. Some key applications requiring optical signal modulation (directly or indirectly) include fast signal keying [3,4], beam steering such as in phased arrays for LIDAR [5,6], and more recently neuromorphic computing by providing the nonlinear activation function of a perceptron [7,8]. Bringing optical modulators into technological context, in recent years both communication and computing have experienced increased challenges to deliver performance, while demand for higher data handling capacity at lower energy-per-bit functions is yet rising. The latter is driven by the rapid increase of internet and data-heavy applications (i.e. media streaming & cloud computing), and is expected to continue to rise with progressing digitalization driven by densification of connected networks and edge-computing known as internet-of-things [9]. It is understood that electronic interconnects limit data bandwidth, while optics enables natural data-handling parallelism given by the bosonic nature of photons and the weak interaction between light and matter, respectively [10-12]. Both communication and emerging computing systems improve performance with increased modulator speed, wide spectral bandwidth, and reduced footprint and energy-per-bit functions [8, 13, 14]. The performance of a modulator, being a three-terminal device, is mainly driven by electrostatics (similar to a transistor), material index change capability, the effective material broadening (state-smearing), and optical mode-engineering. We recently showed that reducing both the mode area and effective broadening of the active (i.e. modulate) material are critical to improve modulation performance [8, 15]. This can be achieved in the former by ensuring a maximum field overlap of the optical mode with the active material by 'squeezing' the optical mode to sub-diffraction limited modes. In the latter (broadening) by trivially cooling the device, increased material quality (i.e. reduced (in)homogeneous broadening), or selecting an active material with a natural lower broadening such as low-dimensional materials. These include predominantly quantum-dots and wells, but also van-der-Waals layered 2-dimensional (2D) materials. We previously found, that the material $Q$-factor is generally lower than that of a cavity ('photonic-$Q$'), thus another design rule for modulators is to avoid material resonances since it introduces loss. An optimized approach, however, is to utilize photonic enhancement such as via cavity or waveguide dispersion engineering. Here we caution high cavity $Q$ approaches, as the long photon lifetime limits the remodulation of the active material; a $Q$ of mid $10^4$, for instance, results in a RC-limited speed of about 10 GHz, which is not considered fast under current signal keying standards. Modulators are classified by as electro-absorptive ($\Delta\kappa$) or electro refractive ($\Delta n$) depending on optical property of the material changed due to the applied electric field. A wide variety of material-index change mechanisms exist to include state-filling, Pauli-Blocking, exciton modulation, or free carriers. Material wise, the contestant to production-level photonic integration modulators are silicon-based devices [16-18]. Electro-optically, silicon shows weak ($\Delta n \sim 10^{-4}$-$10^{-3}$) free-carrier based dispersion [19]. As a result, silicon modulators are bulky (~mm long), require relatively high voltage levels (2-5V), and have a relatively high insertion loss (~3-5dB) given the long length. Here, we consider graphene as an active material for modulation due to its electro-optic properties originating from its unique band structure where charge carriers are governed by the relativistic Dirac equation [20,21]. The high mobility of graphene allows reducing the RC-delay time via a lowered contact resistance [8, 22-24]. Graphene has the potential to fulfill the current state of the art modulator performance due to following special features; firstly, it exhibits the highest charge carrier mobility (~$10^5$ cm$^2$/V-s) at room temperature since the electrons in graphene behave as massless Dirac fermions, which is essential to obtain faster modulation (> 10's GHz) [22, 25]. Secondly, due to its linear band structure, it can absorb the electromagnetic wave over a spectrally broad wavelength band, i.e. from UV to Terahertz even in micro-wave region gives the opportunities for operating at Telecom windows [26]. Thirdly, graphene exhibits strong light matter interaction (~2.3% absorption) in spite of being a single atom thick material (~0.34 nm) which can offer exceptionally high non-linearity for ultrafast light [27, 28]. Fourthly, the interband transition of graphene can be controlled by tuning the Fermi-level ($E_F$) by means of electrical gating, optical excitation, or chemical doping [29-31]. Here, graphene becomes opaque when the Fermi level falls outside of $\pm h\nu_0/2$ due to the Pauli blocking; otherwise graphene is transparent for a photon energy up to $2E_F$ [32]. Fifth, graphene additionally fortifies 2D plasmons due to its free carrier absorption (i.e. intraband transition), which exhibits strong optical confinement in infrared wavelength region [33, 34]. Finally, graphene can be integrated on any other substrate and selectively on photonic devices such as waveguides, fibers, optical micro cavities, and most importantly with mature Si CMOS platform due to the mechanical flexibility and chemical robustness [35-37]. In this work, we assess the state of the art of graphene-based electro-optic modulators. Unlike simply providing a collection-summary of published work, we give a synthesized analysis starting from a device evolution benchmark, over graphene optical models, waveguide mode considerations, to recent device demonstrations showcasing the state-of-the-art.

## 2. EOM Performance Matrix

Electro-optic modulators being key building block for photonic integrated circuit (PIC), have attracted attention in recent years due to their ability to provide signal encoding. As aforementioned, the important performance matrix for EOMs include modulation speed, power consumption, and (less importantly) the footprint area (possibly 3D volume). In this section, we summarize selected studies on EOM development focusing on active material selection and photonic platform design using a unified performance benchmark, where we define the figure-of-merit (FOM) as $FOM = \frac{Data\ capacity}{Power \times Cost} = \frac{f_{3dB}}{(E/bit) \times Area}$ [16]. For seamless integration with CMOS technology, extensive investigation has been made to utilize the classic semiconductor, silicon, for electro-optic modulation by using free carrier injection/depletion. However, due to the intrinsically low carrier concentration, and weak free carrier dispersion of silicon, either a long arm Mach-Zehnder interferometer (MZI) or a low loss, high-$Q$ resonators are needed to increase the light-matter interaction (LMI). Since the RC-delay of the device and power consumption both suffer from larger device volume, both pioneering work on silicon MZI and micro-ring modulators have a rather low overall FOM, which sets our baseline for the performance matrix (Table and Fig. 1) [17, 18]. With the breakthrough on low dimensional materials, graphene has shown great tunability as an absorptive material, as demonstrated firstly in 2011 by integrating this atomic thin material onto conventional photonic silicon waveguide [26]. The silicon waveguide used has is rather compact (tens of micrometer) benefiting from the drastic change of graphene's absorption compared to silicon. However, although the waveguide dimension was optimized to maximize the mode overlap between the propagating mode and the active material (graphene), at telecom wavelength, the conventional photonic mode still has a certain cutoff physical dimension (220 nm for Si at 1550 nm wavelength), which is usually much larger compare to the 0.34 nm single atom layer active material graphene. Thus, for either graphene EOM work from 2011 or 2012 in Table 1 the device dimension is still bulky due to the low LMI [26, 38]. Nevertheless, in 2015 the Lipson-group demonstrated a graphene modulator with 30 GHz-fast device using a low-loss $Si_3N_4$ micro-ring resonator as the underlying photonic platform. Yet this device still suffers from the bulky dimension and reduced thermal stability from the ring resonator [25].

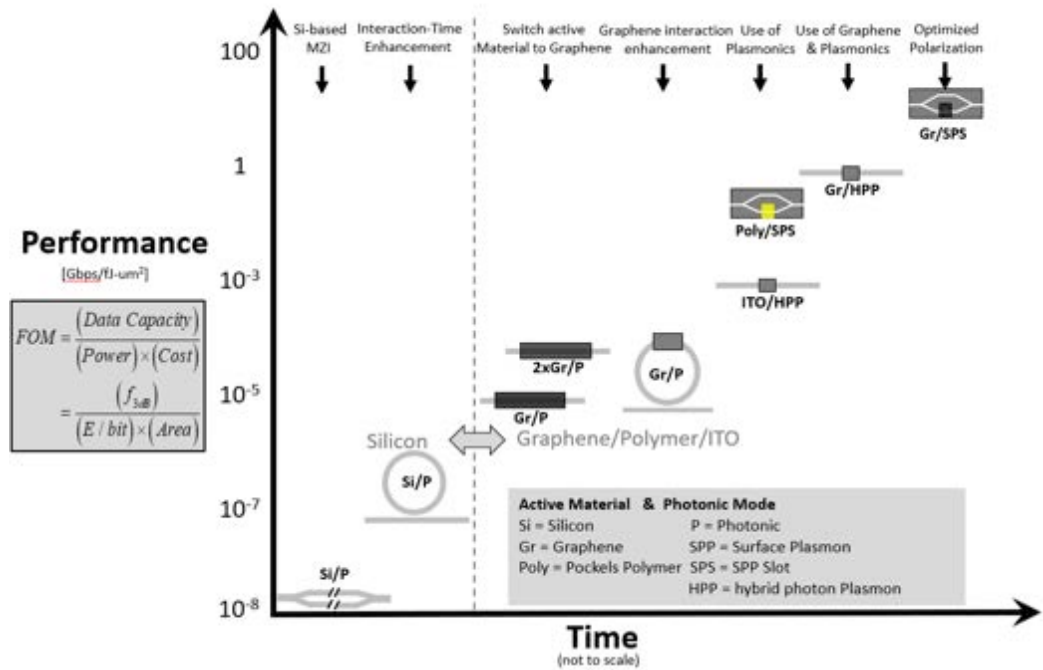

**Figure 1.** The roadmap of EOM development using the pre-defined performance matrix. Graphene showed predominant advantage compared to silicon as the active material selection, while plasmonic mode (HPP/SPP/SPS) provide a much higher light-active material interaction by squeezing the mode into the active material region. Thus the integration of unity index change material and plasmonic mode could reduce the device size and bring the performance to the upper side sweet spot.

On the other hand, increased optical confinement has been explored such as in plasmonic modulators [39-41]. Here, the plasmonic modes increases the modal overlap factor of the active material with that of the optical mode, and provides a free metallic contact which has a smaller series resistance in the device, hence reducing the RC delay compared to the photonic platform which usually requires doped semiconductors as the electrical contact. By squeezing the mode into the active material region to increase the LMI and thus the FOM, modulators have been demonstrated by using other active materials such as transparent conductive oxides (e.g. ITO), but also Pockels-effect field-driven (non charge-drive) materials such as polymers [39-41]. Following this hybrid0-integration scheme (e.g. silicon plus novel-material) Leuthold's group has shown that the latter enables a platform for ultra-fast modulation approaching 200 GHz. Lately, benefiting from the platform design and material selection, studies on integrating graphene with plasmonic modes have shown promising result, as the FOM increases [42, 43]. This is enabled by a) reducing the optical mode area via high plasmonic mode confinement from slot-like waveguide modes, b) the use of unity-strong tunable index materials such as graphene, c) an improved electrostatics by using thin high-k dielectrics, and d) and low-contact resistances supported by metallic synergistic contacts being part of the design. In the following sections, we will review these designs and compare different plasmonic modes for graphene integration. With optimized designs in mind, we discuss how the modulator FOM can be further improved (Fig. 1).

| Year | Mechanism | Bandwidth (GHz) | Footprint (um$^2$) | E/bit (fJ) | FOM | Ref |
|---|---|---|---|---|---|---|
| 2004 | Si MZI | 3 | 10000 | 50000 | 6E-9 | [18] |
| 2005 | SI Ring | 1.5 | 150 | 20000 | 5E-7 | [17] |
| 2011 | Graphene/P | 1 | 40 | 1000 | 2.5E-6 | [26] |
| 2012 | 2x Graphene/P | 10 | 20 | 2000 | 2.5E-5 | [38] |
| 2012 | HPP ITO | 25 | 5 | 56 | 8.9E-2 | [39, 40] |
| 2015 | Poly/SPS | 72 | 20 | 25 | 1.4E-1 | [41] |
| 2015 | Graphene Ring | 30 | 6400 | 80 | 5.8E-5 | [25] |
| 2017 | Graphene HPP | 20* | 20 | 3 | 3.3E-1 | [42] |
| 2017 | Graphene SPS * | 100 | 3 | 0.4 | 83 | [43] |

**Table 1.** The performance comparison table for highlighted EOM works.

## 3. Different Models for Graphene's Optical Properties

While several graphene conductivity models are being discussed in the literature including the Random-Phase-Approximation [44], the Kubo formulism is also being used [45]. We note that the discussion about the relative contributions of free-carriers (intraband) versus the band-to-band (interband) transitions is still ongoing in the field.

### 3.1. Kubo Model and Different Approximations

Graphene shows anisotropic material properties given its dimensions: in its honeycomb like lattice plane, the in-plane permittivity ($\epsilon_{||}$) can be tuned by varying its chemical potential $\mu_c$, whereas the out-of-plane permittivity is reported to remain constant around 2.5, resembling that of graphite [46]. Here, we model graphene by a surface conductivity model $\sigma(\omega, \mu_c, \gamma, T)$, where $\omega$ is the angular frequency, $\mu_c$ is the chemical potential, $\gamma$ is the scattering rate and $T$ is the temperature. The Kubo formula describing the conductivity is given by Equation. (2),

$$\sigma(\omega,\mu,\gamma,T) = \frac{iq^2(\omega - i2\gamma)}{\pi\hbar^2}\left[\frac{1}{(\omega - i2\gamma)^2}\int_0^\infty E\left(\frac{\partial f_d(E)}{\partial E} - \frac{\partial f_d(-E)}{\partial E}\right)dE - \int_0^\infty \frac{f_d(-E) - f_d(E)}{(\omega - i2\gamma)^2 - 4(E/\hbar)^2}dE\right] \quad (1)$$

where $f_d(E) = 1/\left[e^{(E-\mu_c)/k_BT} + 1\right]$ is the Fermi-Dirac distribution function and $\hbar$ is the reduced Planck's constant $\hbar = h/2\pi$, $\gamma$ is taken as 0.001 eV. We can calculate the in-plane conductivity $\sigma_{||}$ from the Kubo formula followed by estimation of the in-plane permittivity as $\epsilon_{||} = 1 - \frac{\sigma_{||}}{i\omega\epsilon_0\Delta}$ where $\Delta = 0.35$ nm, is the thickness of a single graphene sheet. The corresponding refractive index and extinction coefficient are found from the complex in-plane permittivity [43]. The first integral in (1) describes intraband motion of free electrons near energy **E** or holes near the energy –**E**, while the second integral is due to momentum-conserving ('vertical') transitions between the states with energies –**E** in the valence and **E** in the conduction band. The intraband contribution is similar to the Drude response in Si and ITO, and, since at optical frequencies in telecom region ω>>γ, the intraband conductivity in this region is mostly imaginary, i.e. their contribution to the dielectric constant is mostly real. In contrast, the interband contribution has a pole in the vicinity of ω=2**E**/ℏ hence it can contribute to both real and imaginary parts of optical conductivity, i.e. to both extinction coefficient and refractive index. Both intra- and interband terms in (1) depend on the temperature, *T*. In the end every carrier-driven modulator will be limited by a voltage-sharpness limited by the carrier-distribution smearing width, which can be estimated by the Boltzmann Approximation to be 3-4$k_BT$ <100meV. However, we note that all experimental EO modulators operate far from this limit, requiring at least 10× of that amount (i.e usually a few Volts).

With applied gate voltage, the chemical potential $\mu_c$ can be tuned, and, according to equation (1), the real part of conductivity (imaginary part of permittivity) at frequencies close to $\omega_0 = |\mu_c|/\hbar$ experiences sharp change. The real part of permittivity also experiences change, mostly within few *γ*'s from $\omega_0$ but as shown in literature, it is the large change of absorption that is the most attractive feature of graphene. It is important to note that only the in-plane component of the electric field experiences large change in absorption, which has consequences for the modal overlap factor, *Γ*. The voltage bias for the metal and silicon are denoted as $V_f$ and $V_{Si}$ respectively. From Maxwell's equations, the carrier density *n* (i.e. ± for e⁻/h⁺) in graphene has to be equal to the change in the normal component of the electric displacement current across graphene, $\epsilon_{ox}\mathcal{E}_{ox} - \epsilon_{Si}\mathcal{E}_{Si} = en$, where $\epsilon_{ox}$ and $\epsilon_{Si}$ is the electric permittivity of the oxide and silicon layer, $\mathcal{E}_{ox}$ is the electric field across the oxide, and $\mathcal{E}_{Si}$ the electric field in the Silicon adjacent to graphene. The Dirac point energy in graphene $E_D$ is related to the carrier density *n* via, $n = \int_{E_D}^\infty f(E)\mathcal{D}(E - E_D)dE \approx sgn(-E_D)\frac{(E_D)^2}{\pi\hbar^2 v_f^2}$, where *f(E)* is the Fermi-Dirac distribution and $\mathcal{D}(E - E_D)$ is the density-of-states in graphene. Notice, the limit of $[E_D] \gg k_BT$ was used. *T* = Temperature, $v_f$ = Fermi velocity of graphene. The electric field across the oxide is given by, $\mathcal{E}_{ox} \approx$

$\frac{1}{et_{ox}}(\Phi_G - \Phi_m + eV_f)$, where $\Phi_G$ and $\Phi_m$ are the workfunctions of graphene ($\Phi_G = 4.6 eV$) and the metal (eg. $\Phi_{Cu} = 4.65 eV$), respectively. With the electron affinity of Silicon, $\chi_{Si} = 4.05 eV$, a Schottky junction exists. Here, we consider a p-doped SOI-based waveguide core with doping $n_A$. The electric field in the Silicon adjacent to graphene is given by, $\mathcal{E}_{Si} = \sqrt{\frac{2en_A(V_{bi}-V_{Si})}{\epsilon_{Si}}}$, where $V_{bi}$ is the built-in potential in Silicon. Solving these equations relates the graphene's chemical potential $\mu_c$, to the applied voltage, $V_f$.

The Kubo model can have various approximations and we have used the Hanson approximation for the interband transitions at 0 K in previous works [15, 47]. Here we want to compare the diversity among different results operating at $\lambda = 1550\ nm$. All of the results depicted show the similar trend in $n$ and $\kappa$ (Fig. 2). $\kappa$ decreases as Pauli blocking takes effect, and starts to grow again around 0.5 eV due to free carrier intraband absorption. Theoretical models for calculating the optical conductivity of graphene differ from each other since different assumptions were made; for instance, Gusynin et al. developed a frequency dependent electrical conductivity tensor using the Kubo formulation [48]. The scattering rate is assumed to be $\Gamma = 17$ K, Fermi velocity in graphene is taken as $v_F = 1.1 \times 10^6$ ms$^{-1}$ [48]. Stauber et al. also developed a method based on the Kubo formulation to calculate the optical conductivity of graphene at 300 K by taking into account its full density of states and found that in the optical regime the corrections to the Dirac cone approximation are quite small, where hopping parameter is assumed as 2.7 eV [49]. The Hanson approach is the one we use in this work which approximates the interband transitions at 0 K, but the intraband transitions take the temperature dependence into account [45]. Ooi et al. also report graphene complex index data based on different approximations for the Kubo model at T = 300 K [50]. The results from Gosciniak et al. are also derived from the Kubo formula [51].

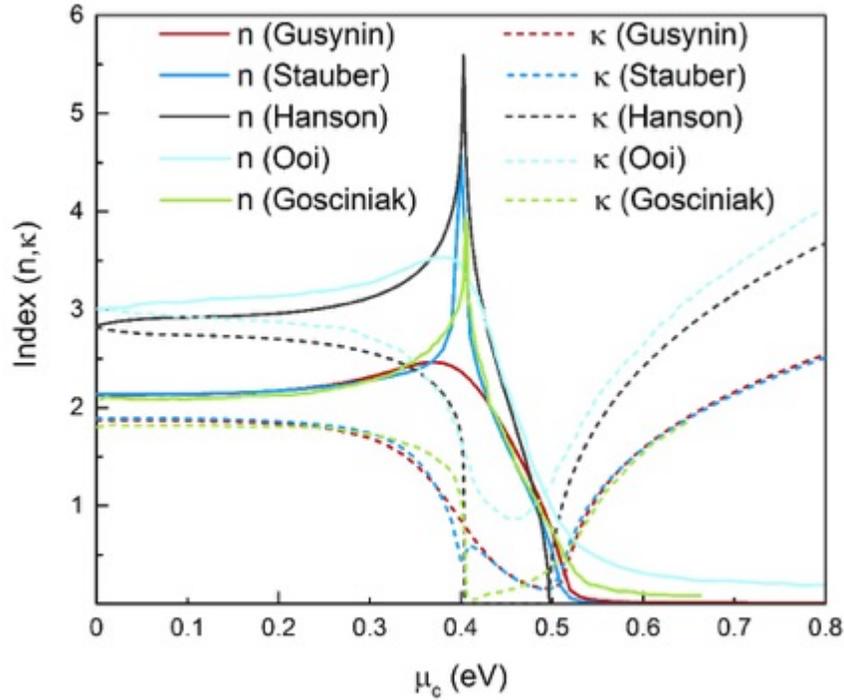

**Figure 2.** Different n and $\kappa$ results for graphene based on Kubo formula at $\lambda = 1550\ nm$. Based on results from Gusynin et al. [48], Stauber et al. (T = 300 K) [49], Hanson (T = 0 K for interband and T = 300 K for intraband transitions) [45], Ooi et al. (T = 300 K) [50], and Gosciniak et al. [51].

All these $n$ and $\kappa$ data reported follow similar trends with $n$ decreasing in the Pauli blocking range and staying at a minimum value for the entire chemical potential range; and $\kappa$ decreases to a minimum near 0.4–0.45 V and grows again due to intraband absorption. All the data sets have n and $\kappa$ intersect each other near 0.5 eV where a transition from dielectric graphene to metallic graphene can be observed. The start points in chemical potential i.e. 0 eV data are different for each result, and as a result all the plots are shifted with similar trends. This can be understood from the fact that many other factors such as quality of the graphene sheet, growth/transfer techniques, impurity, trap states, measurement conditions/criteria etc impact the complex graphene conductivity. Some other notable approaches towards characterizing graphene optical properties, beyond those in Fig. 1, are also available and can be utilized. Peres et al. studied the electronic properties of graphene in the presence of defects, and electron-electron interaction as a function of temperature, external frequency, gate voltage, and magnetic field [52]. Wunsch et al. and Hwang and Das Sarma developed similar methods to calculate the conductivity of graphene from the polarization based on the Dirac cone approximation for a finite chemical potential and arbitrary radian frequency, respectively [53, 54]. Simsek generated a closed-form approximate expression based on Stauber's method, which generates results with less than 0.8% maximum absolute error for $\lambda > 250$ nm [55]. The photons with energies larger than $2\mu_c$ experience absorption, while the photons with smaller energies do not, therefore, when the chemical potential moves across $\mu_c = \hbar\omega/2 \approx 0.4 eV$ the extinction coefficient undergoes rapid downward change as evidenced by the sharp negative peak of $\delta\kappa/\delta\mu_c \sim -\delta(\mu_c - 0.4)$, where $\delta(x)$ is the delta function. According to Kramers-Kronig transform the real part of the refractive index also experiences the rapid change in the vicinity of $\mu_c \approx 0.4\ eV$ as $\delta n/\delta\mu_c \sim -1/(\mu_c - 0.4)$. In practice, a large change in the extinction coefficient occurs in the range of $\mu_c$ values near 0.35–0.45 eV as the Pauli blocking effect can be smeared for ~0.1 eV at room temperature. The modulation technique we investigate here involves interband transitions, whereas intraband free carrier absorption can rise even after the Pauli blocking range of ~0.1 eV and can be approximated by the Drude model. Depending on these interband transitions of Pauli blocking and temperature-dependent smearing of the transition, $n$–dominance is present for $\mu_c$ ranges past the transition until the free carrier absorptions start around 0.5 eV due to very low $\kappa$ values there and abruptly changing $n$ values. $\kappa$–dominance is on either side of the $n$–dominant region.

### 3.2. RPA Model

Graphene optical response is modeled using local random phase approximation (local RPA) in literature [44]. At temperature T, the 2D conductivity of graphene is given by [56]

$$\sigma_{RPA}(\omega) = \frac{2e^2 k_B T}{\pi \hbar^2} \frac{i}{\omega + i/\tau} \ln\left|2\cosh\left(\frac{\mu}{2k_B T}\right)\right| + \frac{e^2}{4\hbar}\left\{H(\omega/2,T) + \frac{4i\omega}{\pi}\int_0^\infty d\zeta\ \frac{H(\zeta,T) - H(\omega/2,T)}{\omega^2 - 4\zeta^2}\right\} \quad (2)$$

where $H(\omega, T) = \sinh(\hbar\omega/k_B T)/\{\cosh(\mu_c/(k_B T)) + \cosh(\hbar\omega/k_B T)\}$. The first term in (2) represents intraband contribution, and the remaining terms are contributions of the interband transitions to the total graphene conductivity. Here, $\tau$ is the electron relaxation time. While Landau damping itself is already included in the conductivity model, the relaxation time typically has other contributions from: a) impurity scattering, b) scattering with phonons ($\hbar\omega_{Ph} = 0.2\ eV$) in graphene and phonon modes of polar substrates, c) higher-order processes such as phonon coupled to e-h excitations (which have to be treated separately), etc [57-59]. In literature, relaxation times as long as 1000 fs have been reported [60, 61]. However, for frequencies larger than the optical phonon frequency of graphene, typically $\tau \approx 50$ fs [57]. In this work, we use a graphene DC mobility $\mu_{DC}$ of $10^4$ cm$^2$V$^{-1}$s$^{-1}$, which is closer to most experimental reported values, but lower than the record-high measurements. The mobility and relaxation time are related by $\tau = \mu_{DC} E_F / e v_F^2$ [62]. The relaxation times considered in this work are in the same order of magnitude range as in [58].

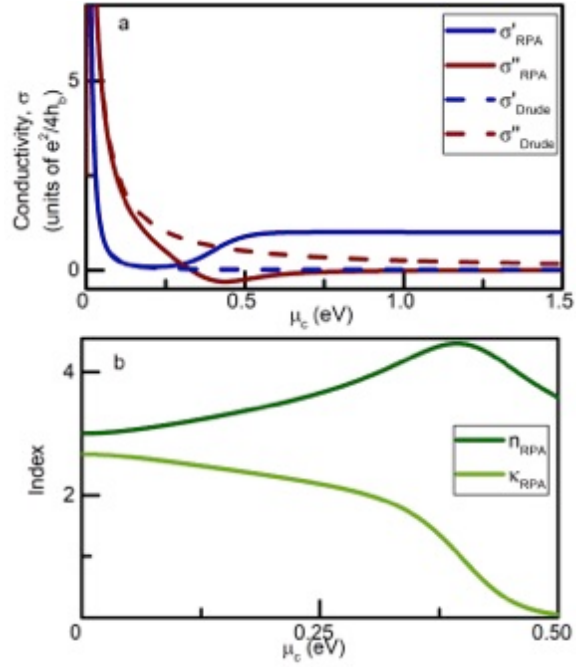

**Figure 3.** Graphene electro-optical properties dictated by the local random phase approximation (RPA) model; **a.** Single layer graphen conductivity, $\sigma$ (real and imaginary parts); and **b.** real and imaginary index, $n$ and $\kappa$ vs. chemical potential, $\mu_c$.

The RPA model closely follows the Drude approximations for the intraband transitions, i.e. free carrier absorption in graphene. But responses from the interband transitions are dissimilar to the Kubo model. Hence, the optical index of graphene, both real and imaginary parts, in the local-RPA model are noticeably different from the values generated by the Kubo model. Since there is this dissimilarity in the optical index results for both the models, and ongoing discussions in the field to articulate accurate results from models, we choose to derive expressions for graphene interband absorption from the first principles in the following section.

### 3.3. Ab-initio approach for Graphene interband absorption

In order for us to derive expressions for graphene absorption relevant to optical modulators, we first define relevant waveguide modal parameters such as effective area and thickness. A generic picture of a current-driven EAM is given in Fig. 4(a), defining the modulator length, $L$ and cross-sectional width, $W$. The active layer (thickness, $d_a$) is sandwiched between two cladding layers for gating. With applied drive voltage, $V_d$ the drive current, $I_d$ flows injecting charges injected into the active layer. Charge, $-Q$ in the active layer and $+Q$ in the gate, are formed by such electrostatic gating. Key to the modulator's performance is a strong LMI, which points towards requiring a small effective mode area, $S_{eff}$, or simply in one dimension (1D), a short effective thickness, $t_{eff}$. To determine the effective area of the waveguide we evaluate the Poynting vector, $S(x) = E_y H_x - E_x H_y$. According to Maxwell equations $\frac{\partial}{\partial z} H_x = \beta H_x = \omega n^2(x)\epsilon_0 E_x$ and $\frac{\partial}{\partial z} H_y = \beta H_y = -\omega n^2(x)\epsilon_0 E_x$. Thus, $S(x) = \frac{\omega n^2(x,y)\epsilon_0 (E_y^2+E_x^2)}{2\beta} = \frac{n^2(x,y)(E_y^2+E_x^2)}{2 n_{eff} \eta_0}$, where the effective index has been introduced as $n_{eff} = \beta c/\omega$. The total power flow is then $P = \iint_{-\infty}^{\infty} S(x,y) dx dy = \frac{1}{2 n_{eff} \eta_0} \iint_{-\infty}^{\infty} n^2(x,y)(E_y^2 + E_x^2) dx dy = \frac{n_{eff} E_{a0}^2}{2\eta_0} S_{eff}$ where $E_{a0}$ is the magnitude of the transverse electric field in the middle of active layer (Fig.1). The effective area of the waveguide is given by

$$S_{eff} = \frac{1}{n_{eff}^2} \iint_{-\infty}^{\infty} n^2(x,y)\left(E_y^2 + E_x^2\right)dxdy / E_{a0}^2 \qquad (3)$$

This definition of the effective area may differ from others in the literature, but the difference is small and involves the distinction between the effective and group indices. We note that this definition includes the fact that the active layer is not necessary at the center of the waveguide, i.e. $E_a$ may not be the peak electric field. Another relevant parameter, the effective thickness of the waveguide is $t_{eff} = S_{eff}/W$.

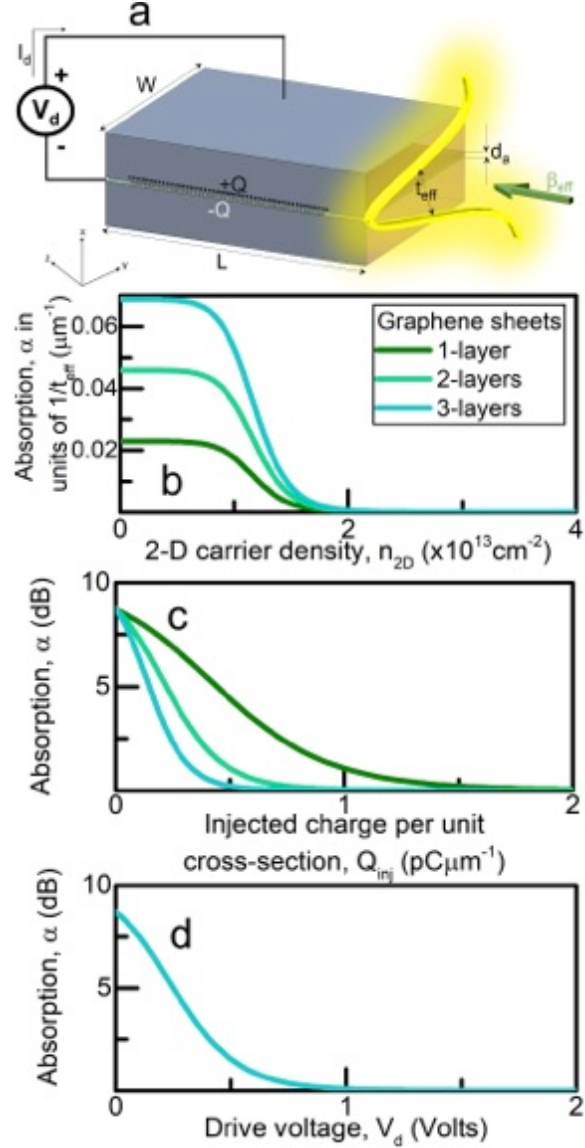

**Figure 4. a.** Schematic of the waveguide structure with a biasing scheme to the active region (drive voltage, $V_d$ and current (charge) flow into the active layer, $I_d$), the charge -Q in the active layer and +Q in the gate is induced. and propagation direction (indicated with green arrow, $\beta_{eff}$). The associated electric field in the y-direction is shown. $d_a$ is the width of the graphene sheet and $t_{eff}$ is the effective thickness. The relevant coordinate system used in this work is also included to accompany the text. **b.** Normalized absorption, $\alpha$ in units of $1/t_{eff}$ (μm$^{-1}$) as a function of carrier concentration, $n_{2D}$ (cm$^{-2}$). Optical absorption vs. **c.** injected charge, $Q'$ (pCμm$^{-2}$); and **d.** drive voltage, $V_d$ (Volts); respectively.

The operating principle of graphene for interband modulation is no different from the quantum well (QW) modulator. Even though there have been many derivations of graphene absorption [48, 49], it is instructive to re-derive the inter-band absorption in graphene from the first principle because it allows us to see how closely related it is to absorption by any other 2D structure whether there are any Dirac electrons involved in it. In fact, when the electrons involved in Pauli blocking are located at roughly 0.4 eV away from the Dirac point (assuming a wavelength of 1550 nm = 0.8 eV), all the peculiarity of graphene becomes irrelevant and as a result its performance as an electro-absorption modulator is conceptually no different from any other semiconductor material. In graphene, the matrix element of momentum between valence and conduction band is $P_{cv,gr} = m_0 v_F$, where $v_F \sim 10^8$ cm/s is the Fermi velocity. The matrix element of Hamiltonian in the $p \cdot A$ gauge then becomes $H_{cv} = \frac{1}{2} e v_F \cdot E/\omega$. The Joint density of states can be found next, as

$$\rho_{2D}(\hbar\omega) = \frac{1}{2\pi} \frac{\omega}{\hbar v_F^2} \tag{4}$$

where we have used the dispersion relation for the transition frequency, $\omega = 2kv_F$ and the fact that, in graphene, one deals with both spin and valley ($K, K'$) degeneracies. Substituting it all into Fermi's golden rule we obtain:

$$\frac{dn_{2D}(y)}{dt} = \frac{2\pi}{\hbar} \frac{e^2}{4} E_{eff}^2 \frac{1}{2} \frac{v_F^2}{\omega^2} \frac{1}{2\pi} \frac{\omega}{\hbar v_F^2} [1 - f_c] = \frac{e^2 E_{eff}^2}{8\hbar^2 \omega} [1 - f_c] \tag{5}$$

where each graphene sheet is populated with $n_{2D}$ carriers that are distributed according to the Fermi-Dirac distribution $f_c(E_c) = \{1 + \exp[(E_c - E_f)/k_B T]\}^{-1}$. Subsequently, we obtain the expression for the graphene interband absorption in the waveguide:

$$\alpha_{gr} = \frac{\pi \alpha_0}{n_{eff} t_{eff}} [1 - f_c] \tag{6}$$

Next, we consider the changes in graphene absorption with density of electrons; since the most change in absorption occurs when the Fermi level approaches $E_F = \hbar\omega/2 \gg k_B T$, the relation between density of electrons and Fermi level is $n_{2D} \approx E_F^2/\pi\hbar^2 v_F^2$ and differentiating it, we obtain $dn_{2D}/dE_F \approx 2E_F/\pi\hbar^2 v_F^2$. Now,

$$\frac{d\alpha_{gr}}{dn_{2D}} = \frac{\partial \alpha_{QW}}{\partial f_c} \frac{\partial f_c}{\partial E_{Fc}} \frac{dE_F}{dn_{2D}} = -\sigma_{gr}(\omega) t_{eff}^{-1} \tag{7}$$

Where the differential absorption cross-section is

$$\sigma_{gr}(\omega) = \frac{\pi^2 \alpha_0}{n_{eff}} \frac{1}{k_B T} \frac{\exp[(\hbar\omega/2 - E_{Fc})/k_B T]}{\{1 + \exp[(\hbar\omega/2 - E_{Fc})/(k_B T)]\}^2} \frac{\hbar^2 v_F^2}{2E_F} \tag{8}$$

This expression obviously peaks at $E_F = \hbar\omega/2$, and for maximum absorption change we obtain

$$\sigma_{gr}(\omega) = \frac{\pi^2 \alpha_0}{n_{eff}} \frac{1}{k_B T} \frac{1}{4} \frac{\hbar^2 v_F^2}{\hbar\omega} \approx 9.7 \times 10^{-17} \, cm^2 \frac{1}{n_{eff} k_B T} \tag{9}$$

Finally, similar to the previous section, in order to account for the fact that the absorption does not change linearly with the Fermi level, and also for the additional broadening, the effective broadening is introduced as $\hbar\gamma_{eff} = ((\hbar\gamma)^2 + (3k_B T)^2)^{1/2}$, and

$$\sigma_{gr}(\omega) = \frac{\pi^2 \alpha_0}{n_a} \frac{1}{\hbar\gamma_{eff}} \frac{1}{4} \frac{\hbar^2 v_F^2}{\hbar\omega} \approx 9.7 \times 10^{-17} \, cm^2 \frac{1}{\hbar\gamma_{eff}} \tag{10}$$

Comparing with results for a QW modulator, we see that for a bandgap of ~0.8 eV the effective mass is about $m_c \approx 0.05 m_0$. Assuming the effective mass ratio parameter $\beta_v = 1$, the results are strikingly similar for graphene and QWs which is easy to understand, since graphene and many typical III-V semiconductors have roughly similar matrix transition elements, and for a given transition energy have similar joint densities of states. Similarly, following (4) and (6),

$$\alpha_{gr} = \frac{\pi \alpha_0}{n_{eff} t_{eff}} \frac{1}{\exp\left[\frac{\hbar v_F \sqrt{\pi n_{2D}} - \frac{\hbar\omega}{2}}{k_B T}\right] + 1} \tag{11}$$

Next, we add the intraband conductivity. When the electric field $\frac{1}{2}Ee^{-i\omega t} + c.c$ is applied according to the second law of Newton, $\frac{d}{dt}\hbar k = -\gamma \hbar k - eE$. The Fermi distribution shifts by $\Delta k = -\frac{eE/2}{\hbar(-i\omega+\gamma)} = \frac{eE/2}{\hbar}\frac{\gamma+i\omega}{\omega^2+\gamma^2}$. To calculate surface current, we assume that the temperature is much smaller than the fermi energy $E_f$, and the sum up over the states becomes $\frac{J}{2} = -e\frac{2\times 2}{4\pi^2}\int_0^\pi v_F cos\theta k_F \Delta k cos\theta d\theta = \frac{e^2}{2\pi\hbar}v_F k_F \frac{E}{2}\frac{\gamma+i\omega}{\omega^2+\gamma^2}$, and we obtain $\sigma = \frac{e^2}{2\pi\hbar}\frac{E_f}{\hbar}\frac{\gamma+i\omega}{\omega^2+\gamma^2}$. The real part of conductivity is $\sigma' = \frac{e^2}{2\pi\hbar}\frac{E_f}{\hbar\gamma_{eff}}$ where $\gamma_{eff} = \frac{\omega^2+\gamma^2}{\gamma}$. The energy dissipation inside graphene layer is $\frac{dU}{dt} = \frac{1}{2}\sigma' E^2 = \sigma' \eta_0 I$. Therefore, the absorption coefficient for intraband absorption is $\sigma_{FCA,gr} = \sigma' \eta_0 = \frac{e^2\eta_0}{2\pi\hbar}\frac{E_f}{\hbar\gamma_{eff}} = 2\alpha_0 \frac{E_f}{\hbar\gamma_{eff}}$. This value is much less than interband absorption of graphene since $E_f \approx \hbar\omega/2$, $\alpha_{FCA,gr} \approx \alpha_0 \gamma/\omega \ll \pi\alpha_0$. Therefore, we can neglect the intraband absorption.

Furthermore, we evaluate the change of total absorption $\alpha_{gr}L$ as a function of the injected charge per unit waveguide cross-section, $Q_{inj} = en_{2D}Lt_{eff}W/S_{eff}t_{eff} = en_{2D}L/t_{eff}$ (Fig. 4c). Only the AC charge $Q_{inj} = e(n_{2D} - n_{2D,0})L/t_{eff}$ is shown for graphene, where $n_{2D,0} = 0.9\times 10^{13} cm^{-2}$ is the electron density that brings the Fermi level within $3k_BT$ from the 0.4eV. We obtain the drive voltage via $V_d = Q_{inj}/C_g$, where $C_g$ is the gate capacitance, showing the absorption vs. drive voltage (Fig. 4d).

## 4. Different Waveguide Modes for Graphene Modulators

We study three different mode structures with the aim to explore modulator-suitable material/mode combinations for both electroabsorption (EA) and electrooptic (EO) modulation mechanisms. The aim is to increase the LMIs towards ultra-compact modulators while preserving a high extinction ratio (ER), i.e. modulation depth, and we consider plasmonics as a spatial mode compression tool towards increasing the LMI and compare two distinct plasmonic modes with a bulk-case for comparison. The two plasmonic modes analyzed are the slot waveguide in a metal-insulator-metal (MIM) configuration [63-65], and a hybrid plasmonic polariton (HPP) design in a metal-insulator-semiconductor (MIS) configuration [66, 67]. All mode structures are chosen on top of a $SiO_2$ substrate, thus providing a leveled playing field. The bulk graphene structure consists of placing a single layer graphene on a gate oxide on top of a Si waveguide, thus forming an electrical capacitor. The Si waveguide was chosen to have a height of 200 nm, which supports the 2nd order TM mode resulting in an improved modal overlap (i.e. in-plane electric field) with the active graphene sheet [26]. The underlying silicon waveguide width is taken as diffraction limited $\lambda/n$. It is important to point out that all gate oxides in this work are fixed with an oxide thickness such that we can compare them in a similar standard (Fig. 5).

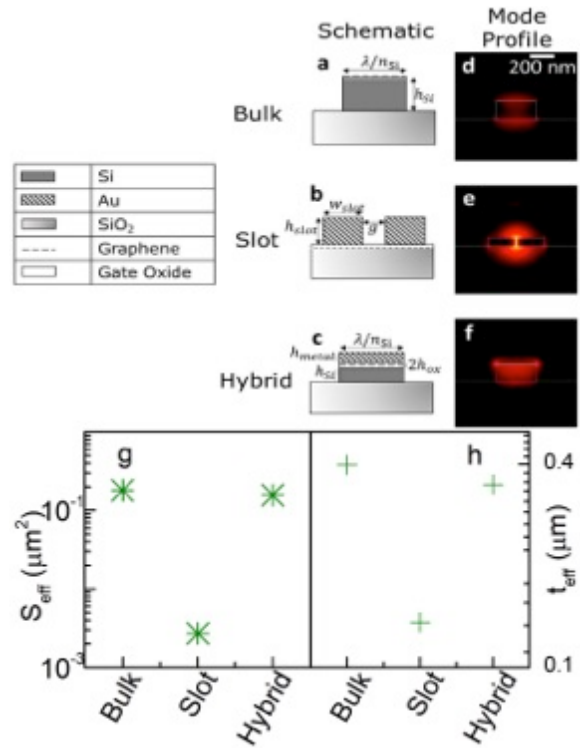

**Figure 5. a-f.** Schematics and FEM simulated mode profiles for different photonic and plasmonic modes for consideration in optical waveguide based modulators. The relevant parameters are $\lambda/n_{Si} = 451\ nm$, $h_{Si} = 200\ nm$, $h_{slot} = 100\ nm$, $w_{slot} = 300\ nm$, $g = 20\ nm$, $h_{metal} = 20\ nm$. All gate oxides have thickness $h_{ox} = 5\ nm$ to ensure similar electrostatics. **g-h.** Effective area and effective thickness for the modes in consideration. The slot graphene mode exhibits both lowest $S_{eff}$ and $t_{eff}$ due to the polariztion matching of the E-field with the active graphene sheet in this mode implying high LMI.

The graphene slot consists of placing a single layer of graphene on top of the SiO₂ substrate separated by a gating oxide of 5 nm. Then two metal pads form the slot structure. Here the gap is 20 nm, which we have found previously to deliver high modulation performance [43]. We note that broader gap dimensions lead to higher order modes, lower optical confinement, and hence lower ER. This value (20 nm) can be understood from two aspects both relating to the fact that metallic confinement beyond 20 nm is not favorable: a) the skin depth of plasmons at telecom wavelengths is about 20-30 nm, and b) the Purcell factor reduces dramatically beyond 10's nm small plasmonic cavities due to high losses and field leakage [14]. Our results indeed confirm a modal confinement to the gap and a high field strength (peak $|E|^2$), which is about 4,500-fold higher compared to the bulk case (Fig. 5e). The hybrid graphene mode is comprised of a metal layer on top of a 10 nm oxide layer, and the graphene single layer is sandwiched inside the oxide. These are stacked on top of a Si waveguide with 200 nm thickness [68]. The hybrid graphene structure shows reasonably high confined modes and a field enhancement of 63 times compared to the bulk case. It is worthy to mention that the slot and HPP modes are comparably lossy without accounting for the material loss to contribute as a byproduct of tuning. As such one intuitively would expect these to be suitable for EAM devices. However, there are also regions where the EOMs (real part index tuning) via phase shifting outperform bulk cases despite the high losses Lastly, only gold (Au) is used for the modal simulations, which has a reasonably low ohmic loss at near infrared wavelengths. However, thermal softness of gold at elevated temperatures (especially for nanostructures) and the incompatibility with CMOS technology, may warrant a future study to explore other plasmonic materials such as $TiO_2$ [69,70].

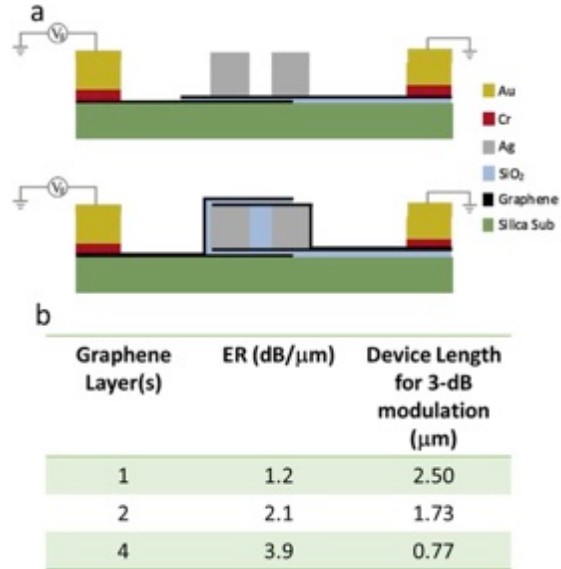

**Figure 6. a.** Multi-layer graphene slot waveguide modulator schematics showing biasing schemes; **b.** Absorption modulator extinction ratio (ER), and physical device lengths needed for achievable 3-dB modulation of the optical signal. Adding more layers demonstrates monotonic improvements in both ER and scaling, but biasing the active layers can become challenging after a few.

Using multiple layers of active material, i.e. multiple graphene sheets, can be an interesting approach to increase the maximum absorption — that will reduce the length, increase voltage and increase speed. We have calculated modulator lengths and ER for up to four layers of actively biased graphene sheets (Fig. 6), beyond a few layers the biasing schemes can become tricky in the slot structure. This multilayer approach is not only limited to the slot structure by default, we just chose to showcase it since a) it performs well among the chosen waveguide designs (Fig. 5), and b) due to graphene's selectivity of interacting only with the in-plane components of the relevant electric field in the mode and the slot's favorable support in achieving such criteria. With graphene, only few layers are enough as the bandgap opens up and we need fewer carriers per layer for desired modulation. Also, bi-layer or multilayer graphene could be used which has a higher density of states (DOS), thus the higher absorption that could be beneficial for EA modulators. The introduced small bandgap is a bi-product of a reduced number of electrons being wallowed around below Fermi level and not contributing to the modulation.

## 5. Device Performance Parameters

Next we combine our details and results from above to analyze exerted modulator device performance metrics. Starting by finding the device length, *L,* we assume an arbitrarily chosen ER = 10dB for the off state absorption, normalized by units of $t_{eff}$ (Fig. 7a). From Fig. 4(c), we can determine the switching charge necessary to obtain this 10dB on-off ratio, based on a simple perturbative estimation, $Q_{sw} \sim 2.2 e/\sigma \sim 10^{-13} - 10^{-12} C/\mu m^2$. Next, we calculate the capacitance, $C_g = \epsilon_0 \epsilon_{gate} L/t_{eff} d_{gate}$ (Fig. 7b). Now, one can finally determine the switching energy-per-bit as $U_{sw} = \frac{1}{2}Q_{sw}V_{sw} = \frac{1}{2}Q_{sw}^2/C_g$ (Fig. 7c). Further, we calculate the 3dB cut off frequency $f_{3-dB} = 1/2\pi R C_g$ assuming $R = 50\Omega$ (Fig. 7d). Such a low resistance may not be realistic for photonic bulk modes, where partial and selective doping has to be used in order to keep both the carrier density and hence optical loss low. In contrast, the metal deployed in plasmonics can also serve as a low resistance contact [8]. As such, the contact resistance may vary by one order of magnitude between photonic bulk designs, and plasmonic waveguide cases. The requirement of a micrometer-tight integration of optoelectronic devices has also been mentioned recently as a viable path for aJ-per-bit efficient devices [8, 71]. The relevant figure of merit (FOM) for modulators, is the ratio of the switching energy and cut-off frequency ('Energy Bandwidth Ratio' or $EBR$), $EBR = U_{sw}/f_{3-dB} = \pi Q_{sw}^2 R$, where evidently lower $EBR$ is desired.

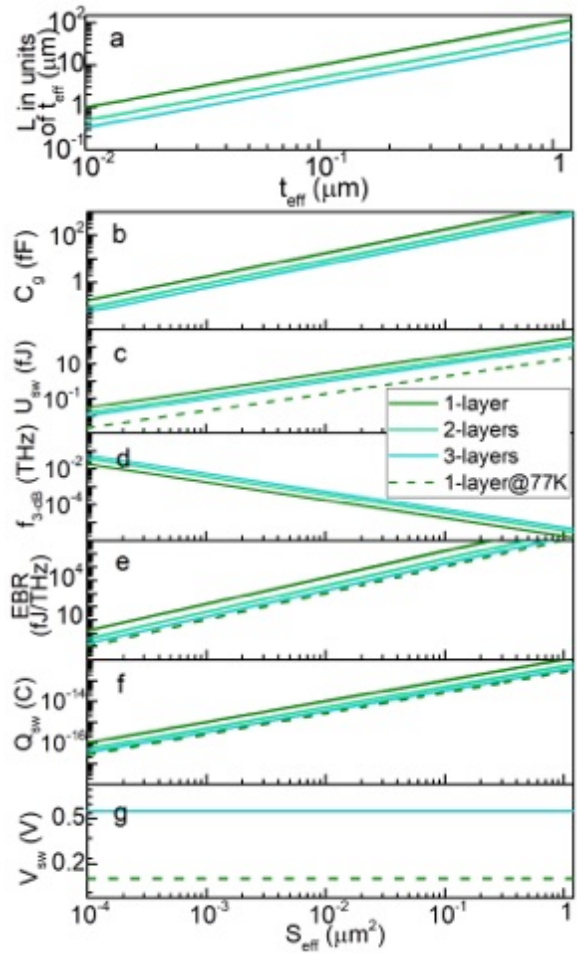

**Figure 7. a.** 10 dB absorption modulator length, $L$ vs. effective thickness, $t_{eff}$ for all the comparable material classes with different broadening. **b-g.** Relevant device parameters vs. effective mode areas, $S_{eff}$ for varying number of graphene layers and thermal cooling at 77K to reduce the effective broadening, $\gamma_{eff}$; **b.** Switching charge, $Q_{SW}$ (C); **c.** Capacitance, $C$ (fF); **d.** Switching energy (E/bit function), $U_{SW}$ (fJ); **e.** 3-dB modulation bandwidth, $f_{3-dB}$ (THz); **f.** Energy-bandwidth ratio, $EBR$ (fJ/THz); and **g.** Switching voltage, $V_{SW}$ (V). The varying amount of broadening corresponding to room temperature, i.e. 300 K for the single layer, and thermal cooling down to 77 K; $d_{gate} = 5$ nm.

To study the dependence on the effective broadening, $\gamma_{eff}$ for QWs, and graphene with change in the operating temperature the device parameters adjust with the scaling factor, $G = T/300$ K as our results are calculated based on the room temperature performances. As such, the switching charge, $Q_{SW}$ scales as $\Gamma$, while the capacitance, $C_g$, stays constant, and so does the 3-dB modulation bandwidth, $f_{3\text{-}dB}$. The modulator length, $L$ is unchanged from the variation in broadening; the switching voltage, $V_{SW}$ scales as $\Gamma$; switching energy, i.e. the energy-per-bit function, $U_{SW}$ and energy-bandwidth ratio, $EBR$ both scale as $\Gamma^2$. Our results demonstrate rationale for cryogenic operation for graphene modulators to cut down energy overhead and achieving better $EBR$. It certainly would be a viable option if operating temperatures would be near cryogenic already, as this does not take into account the energy overhead needed to cool the entire system down to near cryogenic temperatures. Evidently, using multilayers of active graphene can help to lessen the switching energy even lower. Our results show indeed atto-Joule efficient graphene modulators are feasible and the slot structure can fetch 10's of aJ efficient switching at room temperature. We have recently experimentally demonstrated a hybrid graphene absorption modulator requiring 100's of aJ per bit, discussed below.

## 6. Experimental Device Results

The modulator shown here uses a hybrid-plasmon-polariton mode [66], which electrically resembles a MOS capacitor with an additional oxide-Graphene sandwiched inside the MOS oxide (Fig. 8a-c). This optical mode enables a high optical concentration factor resulting in an improved overlap area with low-dimensional materials such as the Graphene. Tuning Graphene's index alters the effective modal index of the waveguide structure. The modal hybridization of a photonic Silicon-on-insulator (SOI) rip mode and a surface plasmon helps to keep the insertion loss from the SOI mode to the modulator switching island, after which, light continues to propagate in silicon. This helps to functionally separate the light manipulation (modulation) from the passive low-loss propagation (in SOI).

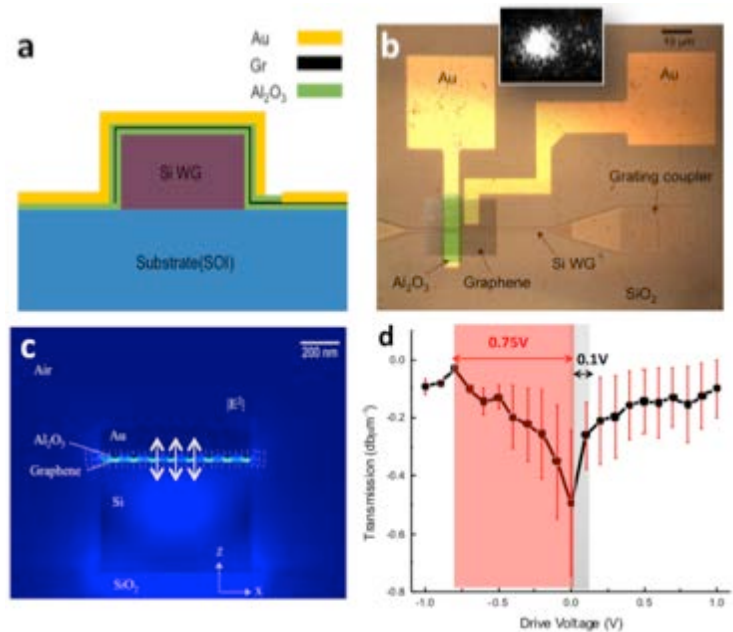

**Figure 8. a.** Cross-sectional schematic of a hybrid-photon-plasmon graphene-based electro-absorption modulator. The modulation mechanism is based on Pauli-blocking upon shifting the Fermi level of graphene via electrostatic gating. **b.** Silicon

waveguide-integrated modulator. A cw laser ($\lambda$ = 1.55 μm) is fiber coupled into the SOI waveguide via grating couplers. Device length, $L$ = 15 and 8 μm, $t_{ox}$ = 5 nm (see text). **c.** Electric field density across the active MOS region of the modulator showing an enhanced field strength coinciding with the active graphene layer. Taking into consideration the polycrystalline grain sizes of the gold from e-beam deposition creates in-plane field vectors inside Graphene. **d.** DC modulator transfer function; normalized modulation depth at different drive voltages ($V_d$). The modulator performance yields a high extinction ratio of 0.1-0.5 dB/μm depending on the bias range. $R_{contact}$ ~ 200Ω.

Probing the modulators transfer function at DC bias, results in steep switching (light ON-OFF-ON) when scanning from negative-zero-positive voltage biases applied to the graphene and the top plasmon-mode metal layer (Fig. 8d). The shape of transfer function originates from graphene's Pauli blocking, thus changing the extinction coefficient, $\kappa$, which is proportional to the real part of the permittivity (Fig. 9a). A charge-driven modulator's switching voltage is given by $V = \frac{Q}{C} \propto \frac{t_{ox}}{\epsilon_{gate}} n_{eff} A_m \frac{\gamma_{eff}}{f_{12}}$, where $Q$ and $C$ are the physical charge and gate capacitance, $t_{ox}$ and $\epsilon_{gate}$ are the gates oxide thickness and permittivity, $n_{eff}$ the effective mode index of the waveguide, $A_m$ the mode area, $f_{12}$ the oscillator strength, $\gamma_{eff}$ the material broadening which is the sum over (in)homogeneous- and temperature broadening (the latter ~ 75meV at room temperature) [72]. Thus, low energy-per-bit functions ($E/bit \sim CV^2$) can be achieved by improving the electrostatics (low $t_{ox}$, high-κ dielectrics), increasing the effective mode index, reducing the mode area (i.e. sub diffraction limited modes), or improving the material quality or temperature reducing broadening and enable a steeper transition across the Fermi-Dirac function. In our device here, we use a thin 5nm oxide of relatively high-κ ($Al_2O_3$, ALD deposited). However, this by itself could not explain our observed switching, which did not match eigenmode simulation results at first. The reason for this is, that the hybrid plasmon mode being TM has a polarization that is perpendicular (z-direction) to the graphene layer (Fig. 8c). This out-of-plane field has vanishing overlap with graphene at the lateral center of the waveguide. This conundrum is lifted when we considered a in-plane field components at the waveguide edge, and b) surface roughness inside the MOS structure near the graphene; measuring the surface roughness of the gold metal deposition showed 10-20 nm large islands consistent with the poly-crystalline films known from e-beam deposition. Adding this to the numerical simulation matched numerical and experimental results well.

In general, the optical interconnect or photonic circuit designer has voltage-choice options; for instance, one could drive the modulator across 0.75V for a 0.4dB/μm steep modulation, resulting in a compact $L$ = 8μm device to achieve an extinction ratio (ER) of 3dB for small signal modulation corresponding to an E/bit of 3.2fJ (red area, Fig. 8d). Alternatively, a bias of only 0.1V or ER = 0.2dB/μm increases the capacitance due to the longer required length $L$ = 15μm, but over-proportionally reduces the energy efficiency resulting in a low 110aJ/bit (grey area, Fig. 8d).

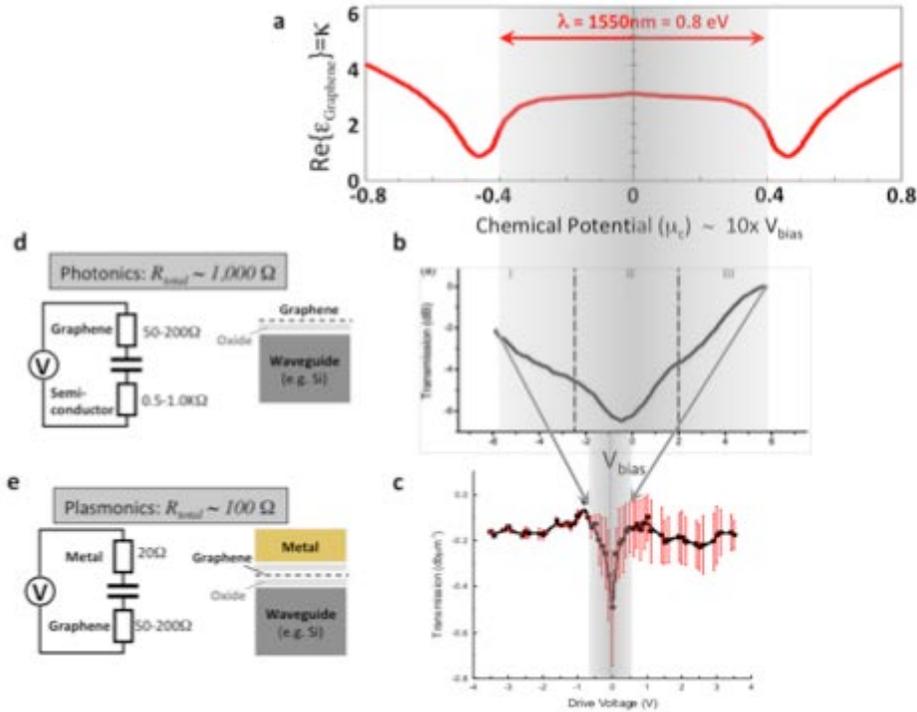

**Figure 9.** Electro-absorption modulator electrostatics and impact of plasmonics. **a,** Graphene's Pauli blocking modulation for extinction coefficient vs. chemical potential (i.e. bias voltage). Kubo formulism used, λ = 1550nm, T = 300K. **b,d** Photonic contact resistance is relatively high (~1,000Ω) to mitigate loss of the long (0.1-1mm) devices. This results in high (5-6) voltage required or modulation [4]. **c,e,** A plasmon mode however, enables 10× lower total contact resistance utilizing the plasmonic metal as a contact. Data from **c** are the same as those in Fig. 8d.

The electrostatic Pauli blocking of graphene requires capacitive gating. An engineering challenge here is to minimize the voltage drop from the contacts to the device region. In photonics (non-plasmonic) devices the weak light-matter-interaction requires up to millimeter-long modulation lengths known from Silicon photonics foundries. This extended length, requires careful loss management, challenging insertion loss (Fig. 9d); typical methods are to highly dope selective regions near (but not including) the active device area. Such bias configuration results in typical contact resistance to the semiconductor terminal of about 100's of ohms. With graphene's contact resistance being in the order of 10-100Ω a photonic device' total contact resistance is about 1,000Ω and requires 5-6 volts of drive voltage [4] (Fig. 9b). In contrast plasmonic modulators are micrometer-compact, which makes their loss-per-length less of a factor. Here, the metal forming the compressed optical mode can be used synergistically as a low-resistance contact right at the active device region. Together, this allows the total resistance of both contacts to be on the order of 100Ω (Fig. 9e). This means, that plasmonics fundamentally allows a higher efficient usage of voltage, which particularly impacts the energy efficiency due to the $\sim V^2$ scaling. We note, that a lower contact resistance will also reduce the power dissipation of monolithically integrated plasmon lasers [11, 73], 2x2 switches [74] and photonic routers [75].

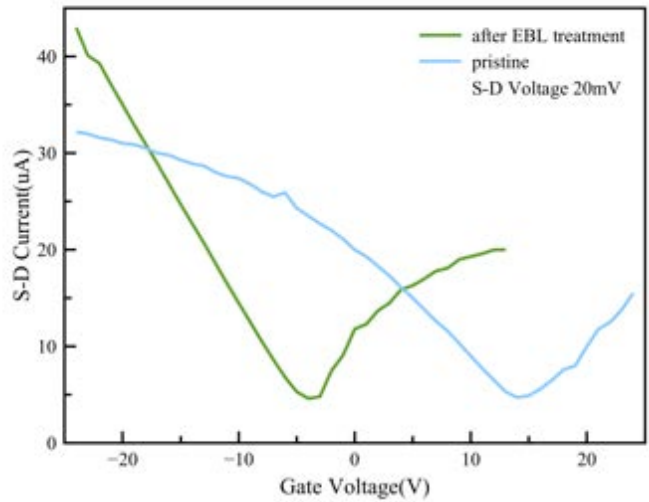

**Figure 10.** Graphene Fermi-level engineering via electron doping in a 3-terminal transistor like measurement (source, drain, gate bias). The Fermi energy of the original (pristine) graphene is 0.96eV and shifts after electron-beam lithography (EBL) treatment to -1.80eV converting graphene from being p-type to n-type. Parameters: $V_{SD}$ = 20mV. Graphene channel dimensions 10μmx10μm. Exposure is done in a dose-controlled method (320μC/cm$^2$) using an EBL system.

Indeed, experimentally graphene's doping level can be tuned via post-transfer processes and probed in a 3-terminal field-effect structure (Fig. 10). The Fermi energy of graphene layer can be altered by metal electrodes deposition on graphene significantly. The graphene's Fermi energy shift upon metal deposition by testing a back-gated field effect transistor design. The graphene layer set on 50nm silicon oxide and the back gate is silicon. The Au contact pads deposited by electron beam deposition. After the deposition of the Au contact pads, the Fermi energy of the graphene channel shifted to -1.80eV, which becomes p-type doped. We found that the electrical properties of graphene can also be shifted back to n-type by exposure to the electron beam. To execute this in a controlled fashion, we selected the writing function from an electron beam system. After the electron beam treatment, we observe an $I$-$V_{gate}$ curve right shift corresponding to a p-type doping. The Fermi energy shifted to about 0.96eV, which changed to n-type doped. This method shows a simple hands-on option to control the Fermi energy of graphene. We note that the ideal chemical potential for a graphene-based modulator may not be at $\mu_c$ = 0eV; that is the steepest index change in the graphene's transfer function (Fig. 3,4,9a) is near 0.4eV (for λ = 1550nm = 0.8eV). Thus, in order to reduce static parasitic power dissipation, it would be elegant to permanently bias graphene to this point without applied bias. The latter could be done either chemically or via a post-processing electron-doping as demonstrated here.

## 7. Summary & outlook

In this work, we discussed all the aspects of graphene modulators starting from the optical properties with various theoretical models, design of different device configuration, fabrication and analysis of their performances. We have developed an ab-initio approach for graphene interband absorption for modulation since different approximations in literature predict ambiguous modulation results. We have shown pathways to optimizing the modulation response of graphene modulators by varying number of active graphene layers in design and varying the operating temperature to achieve sharper transitions, i.e.

reduced broadening. We also predict device performance based on our theoretical model and follow up with experimental device results closely following the perturbative theoretical predictions. We show the pathways to achieving attojoule efficient switching in graphene using plasmonic modes, both theoretically and experimentally. However, there is lot of challenges still remain to be solved in terms of modulation speed, insertion loss and energy consumption to use them as optical interconnect for future data communication and computing technology. One of the main limiting factor of modulation bandwidth is value of RC constant which mainly originates due to the high contact resistance of graphene devices. Several other 2D materials such as transitional metal dichalcogenides (TMDs) and black phosphorus emerged out to be more optically active beyond graphene. The interesting fact about TMDs is that the optical transition is governed by the excitons, where the exciton binding energy can be very high (~few hundreds of meV). Like graphene, the refractive index of TMDs monolayers also can be tuned around exciton resonances by using (CMOS) compatible electrical gating. Henceforth, the large variety of available 2D materials and their heterostructures may outperform the current competing technologies, which could be based on some new modulation mechanism.

**Acknowledgement**
V.S. is supported by ARO (W911NF-16-2-0194), and by AFOSR (FA9550-17-1-0377). H.D. is supported by AFOSR (FA9550-17-P-0014) of the small business innovation research (SBIR) program.